\pdfoutput=1
\documentclass{PoS}

\usepackage[utf8]{inputenc}

\usepackage{amsmath}
\usepackage{mathtools}
\usepackage[capitalise]{cleveref}
\usepackage{tabularx}

\usepackage{wrapfig}
\usepackage[font=footnotesize,labelfont=bf]{caption}
\usepackage[font=footnotesize,labelfont=bf]{subcaption}

\DeclareCaptionSubType*[alph]{table}
\usepackage[wide,ragged]{sidecap}
\sidecaptionvpos{figure}{c}
\sidecaptionvpos{table}{c}

\newcommand{\pt}{\mathrm{pt}}

\usepackage[nomessages]{fp}
\newcommand{\relfigwidth}{0.6}
\FPeval{\sidecaptionrelwidth}{1/\relfigwidth-1}

\clearpage{}

\newcommand{\mpi}{M_\pi}
\newcommand{\mphyspi}{M_{\pi}^\mathrm{phys}}

\newcommand{\fm}{~\mathrm{fm}}

\clearpage{}

\title{Pion structure from twisted mass lattice QCD down to the physical pion mass}

\ShortTitle{tmLQCD pion structure}

\author{\speaker{Bartosz~Kostrzewa}${^\mathrm{a}}$, Maximilian~Oehm${^\mathrm{a}}$, Francesco~Sanfilippo${^\mathrm{b}}$,
        Silvano~Simula${^\mathrm{b}}$, Carsten~Urbach${^\mathrm{a}}$\\
        ${^\mathrm{a}}$ HISKP (Theory), Rheinische Friedrich-Wilhelms Universit{\"a}t Bonn, 53115 Bonn, Germany\\
        ${^\mathrm{b}}$ INFN, Sezione di Roma Tre, Via  della  Vasca  Navale  84, I-00146 Rome, Italy
        
        E-mail: \email{bartosz.kostrzewa@hiskp.uni-bonn.de}}

\abstract{We present an investigation of pion structure based on ETMC $N_f=2$ and $N_f=2+1+1$ twisted mass configurations at maximal twist.
We compute the first moment of the quark momentum fraction of the pion, $\langle x \rangle$ and the electromagnetic form factor, $F_\pi(Q^2)$.
For the latter, momentum is injected using twisted boundary conditions and the calculation is carried out directly at the physical pion mass.
We find that our data are consistent with vector meson dominance and experimental data in the region of small momentum transfer.
For $\langle x \rangle$, we find that our chirally and continuum extrapolated result is compatible with its phenomenological value.\\
\begin{minipage}{\textwidth} \vspace{2cm} \centering \includegraphics[width=0.3\textwidth]{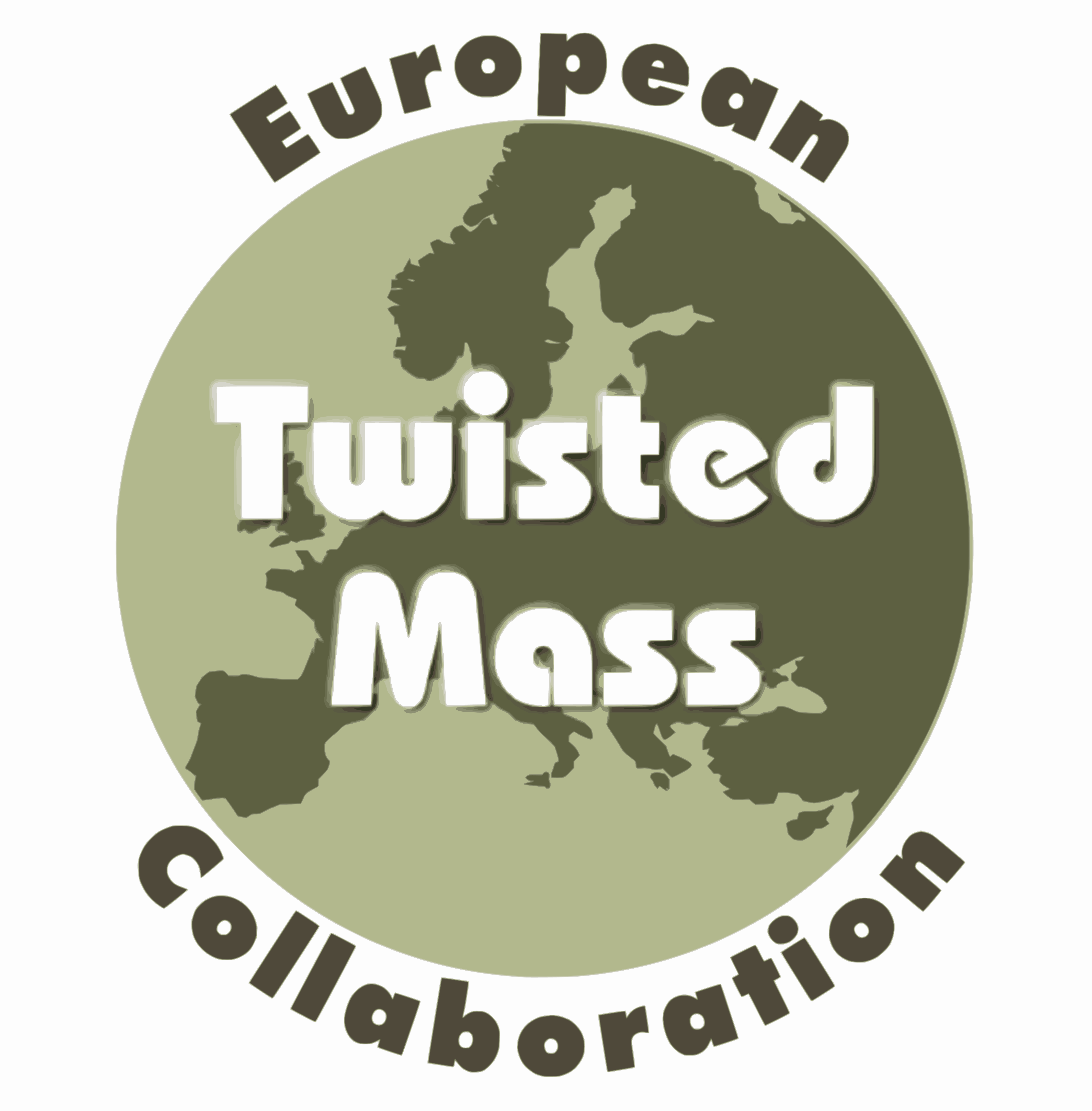} \end{minipage} }

\FullConference{34th annual International Symposium on Lattice Field Theory\\
		24-30 July 2016\\
		University of Southampton, UK}

\begin{document}

\section{Pion Electromagnetic Form Factor} \label{sec:pionff}

The electromagnetic (vector) form factor of the pion, $F_\pi$, parametrizes how the pion deviates from a point particle when probed electromagnetically, thus giving insight on the distribution of its charged constituents.
In addition, its determination in lattice QCD provides an excellent opportunity for the study of chiral logarithms because it depends significantly on the light quark mass.
This also means that a controlled extrapolation to the physical point is a delicate endeavour, such that one would ideally like to perform the computation directly at the physical pion mass.
In the following, we present such a calculation using twisted mass quarks \cite{Frezzotti:2003ni} at maximal twist.
Examples of previous studies in lattice QCD can be found in \cite{Frezzotti:2008dr,Boyle:2008yd,Brandt:2013dua} and a recent computation at the physical pion mass is given in \cite{Koponen:2015tkr}.
Phenomenological determinations are given, for example, in \cite{Blok:2008jy,Huber:2008id,Amendolia:1986wj}.

The form factor is computed from matrix elements of the vector current
\begin{equation*}
  V_\mu(x) = \frac{2}{3} \bar{u}(x) \gamma_\mu u(x) - \frac{1}{3} \bar{d}(x) \gamma_\mu d(x)
\end{equation*}
between pion states.
This gives
\begin{equation*}
  \left\langle \pi^+(\vec{p}') | V_\mu(0) | \pi^+( \vec{p}) \right\rangle
    = ( p_\mu' + p_\mu ) \, F_\pi(Q^2) \, ,
\end{equation*}
where the momentum transfer $q_\mu=( p_\mu - p_\mu' )$ and $Q^2=-q^2$.
In the isospin-symmetric limit, it is sufficient to compute only the $\bar{u}(x)\gamma_\mu u(x)$ insertion, as the two contributions differ only in their sign.

\subsection{Euclidean Correlation Functions and Lattice Setup}

\begin{wrapfigure}{r}{0.4\textwidth}
  \includegraphics[width=\linewidth]{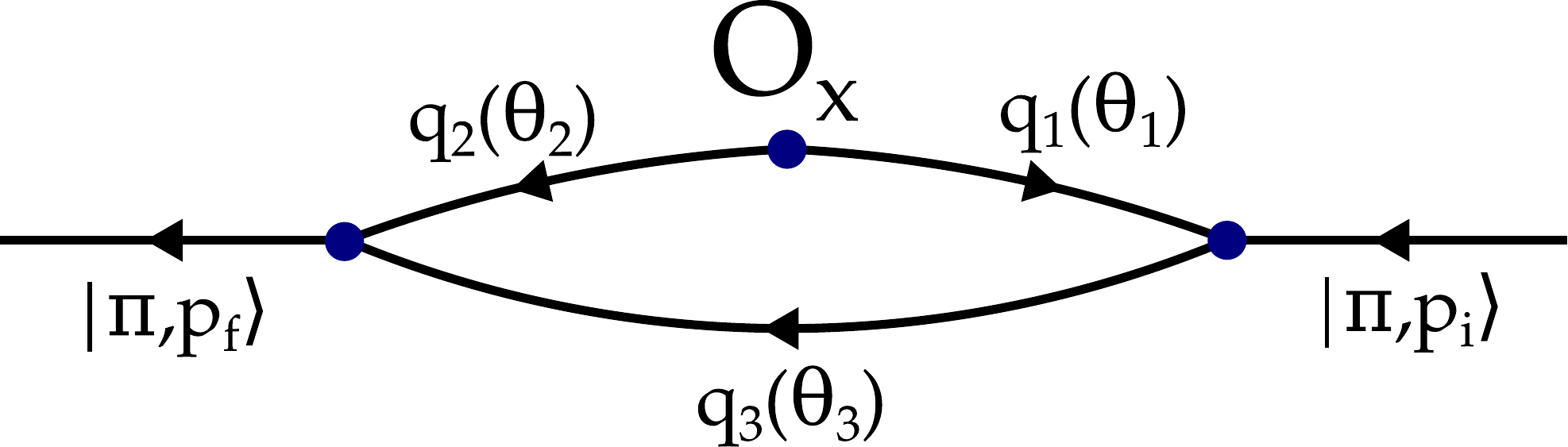}
  \caption{Twisted boundary conditions in the pion three-point function for arbitrary momentum transfer.}
  \label{fig:pionthreep}
\end{wrapfigure}

Working in Euclidean space, we can access the region of space-like momentum transfer, $Q^2>0$, by evaluating ratios of pion two-point functions and three-point functions with the vector current insertion.
To inject arbitrary momenta, we make use of \emph{twisted} boundary conditions (TBCs) \cite{deDivitiis:2004kq,Sachrajda:2004mi,Flynn:2005qn} on the quark fields.
Enforcing $\psi(x+\vec{e}_i L) = e^{i\theta_i} \psi(x)$ on the quark field $\psi$, changes the momentum quantisation condition in finite volume to $p_i = \frac{\theta_i}{L} + \frac{2\pi n_i}{L}$.
This is shown in \cref{fig:pionthreep} for the pion three-point function with independent values of the twist angle for the three quark lines.

For the case of twisted mass quarks, this setup was first studied in \cite{Frezzotti:2008dr} in the Breit frame ($\vec{p}' = -\vec{p}$), which results in a momentum transfer independent of the pion mass
\begin{equation*}
  q^2 = \left[ E_\pi(\vec{p}) - E_\pi(\vec{p}') \right]^2 - 
        \left| \vec{p} - \vec{p}' \right|^2 = -4\left| \vec{p} \right|^2 \, .
\end{equation*}
To obtain Breit frame kinematics with TBCs, we set $\theta_1 = -\theta_2$ and $\theta_3=0$. 

As detailed in \cite{Frezzotti:2008dr}, the required correlation functions can be evaluated efficiently through the usage of the so-called one-end-trick combined with spatial all-to-all propagators from stochastic time-slice sources and the sequential propagator method for the insertion.

With the spatial matrix elements of the vector current vanishing in the Breit frame, we have to compute the following correlation functions
\begin{align*}
  C^{2\pt}(t,\vec{p}) & = \sum_{x,z} \left\langle O_\pi(x) O^\dagger_\pi(z) \right\rangle \delta_{t,t_x-t_z} e^{-i\vec{p} (\vec{x}-\vec{z})} \\
  C^{3\pt}_0(t,t',\vec{p},\vec{p}') & = \sum_{x,y,z} \left\langle O_\pi(y) V_0(x) O^\dagger_\pi(z) \right\rangle
                \delta_{t,t_x-t_z} \delta_{t',t_y-t_z} e^{-i \vec{p}(\vec{x}-\vec{z}) + i\vec{p}'(\vec{x}-\vec{y}) } \, ,
\end{align*}
where $O_\pi(x)=\bar{d}(x)\gamma_5 u(x)$ is the interpolating operator annihilating the $\pi^+$ and $t'=t_{\mathrm{sink}}-t_{\mathrm{source}}$.

Taking the appropriate limits with $T$ being the time extent of the lattice, one obtains 
\begin{align*}
  \lim_{\substack{t\to\infty \\ T \to \infty}} C^{2\pt}(t,\vec{p}) & \to \frac{G_\pi}{2E_\pi(\vec{p})} e^{-E_\pi(\vec{p})t} \\
  \lim_{\substack{t\to\infty \\ (t'-t)\to\infty \\ T \to \infty}} C^{3\pt}_0(t,t',\vec{p},\vec{p}') & \to \frac{G_\pi}{2E_\pi(\vec{p})2E_\pi(\vec{p}')}
        \left\langle \pi^+(\vec{p}') | V_0 | \pi^+(\vec{p}) \right\rangle e^{-E_\pi(\vec{p})t} e^{-E_\pi(\vec{p}')(t'-t)} \, ,
\end{align*}
where $G_\pi$ is the amplitude of the correlation function.
This allows one to construct the ratio
\begin{equation*}
  \lim_{\substack{t\to\infty \\ (t'-t)\to\infty \\ T\to\infty}} \frac{C^{3\pt}_0(t,t',\vec{p},-\vec{p})}{C^{2\pt}(t',\vec{p})} \to 
        \frac{\left\langle \pi^+(-\vec{p}) | V_0 | \pi^+(\vec{p}) \right\rangle}{2E_\pi(\vec{p})} = \frac{1}{Z_V} F_\pi(q^2) \, .
\end{equation*}
To extract $F_\pi(q^2)$, we compute the renormalisation constant of the vector current, $Z_V$, from the ratio of the two and three-point functions at zero momentum transfer and the known normalisation $F_\pi(0)=1$, which implies
\begin{equation*}
  \lim_{\substack{t \to \infty \\ (t' - t) \to \infty \\ T \to \infty}} \, \frac{C^{2\pt}(t,\vec{0})}{C^{3\pt}_0(t,t',\vec{0},\vec{0})} \to Z_V \, .
\end{equation*}

In order to minimize excited state effects while enabling the averaging of forward and backward correlation functions, the source-sink separation is fixed to $t'=T/2$.
On each gauge configuration, multiple source time slices are chosen randomly across the whole time extent, which has been shown to decorrelate measurements from different gauge configurations.

For higher values of the pion momentum, the raw data for the two-point function at $t'=T/2$ can be very noisy, introducing unnecessary variance into the extraction of $F_\pi(q^2)$.
In order to improve our signal, we instead fit the two-point function in an appropriate fit interval and reconstruct it at $t'=T/2$ from the fit parameters, assuming a functional form 
\begin{equation*} 
  C^{2\pt}(t,\vec{p})=\frac{G_\pi}{2E_\pi(\vec{p})} \left[ e^{-E_\pi(\vec{p})t} + e^{-E_\pi(\vec{p})(T-t)} \right] \, ,
\end{equation*}
in a region dominated by the pion and taking into account the effect of propagation around the torus.

\subsection{Results}

The ensembles of gauge configurations for this calculation are introduced in \cite{nf2physpointsimul:2015} and feature charged pion masses $M_\pi \approx [135,250,340] \, \mathrm{MeV}$ and spatial lattice extents between $2.2$ and $5.8$ fm.
\begin{wraptable}{l}{0.46\textwidth}
  \centering
  \footnotesize
  \begin{tabular*}{0.95\linewidth}{@{\extracolsep{\fill}}cccc}
    \hline \hline \\[-2.0ex]
    $M_\pi$ [MeV]      & $L/a$           & $\langle r^2 \rangle~[\mathrm{fm}^2]$ & $c~[\mathrm{fm}^4]$                \\[0.6ex]
    \hline \\[-2.2ex]
      $\approx 340$                    & $24$            & $0.251(11)$                    & $0.0008(01)$                 \\
      $\approx 340$                    & $32$            & $0.270(08)$                    & $0.0014(02)$                 \\
      $\approx250$                    & $24$            & $0.211(22)$                    & $0.0007(02)$                 \\
      $\approx135$                    & $48$            & $0.330(30)$                    & $0.0020(23)$                \\
      $\approx135$                    & $64$            & $0.437(45)$                    & --                          \\
      \hline\hline \\[-2.0ex]
  \end{tabular*}
  \caption{Results of the fit of \cref{eq:F_pi_qsqr_expansion} to the data as shown in \cref{fig:etmc_pion_vectorff_gev}.}
  \label{tab:F_pi_r_c_results}
\end{wraptable}
On each ensemble, we analysed around 250 gauge configurations and computed propagators with between 6 and 15 values of the squared momentum transfer up to $Q^2 \approx 0.22\, \mathrm{GeV}^2$.
For the largest lattice size at the physical pion mass, we currently only use data at one non-zero value of the momentum transfer.

Our determinations of $F_\pi(Q^2)$ for the different ensembles are shown in \cref{fig:etmc_pion_vectorff_gev} together with experimental data for this quantity from \cite{Amendolia:1986wj}.
The lines in the figure are fits of the form
\begin{equation}
  F_\pi(Q^2) = 1 - \frac{\langle r^2 \rangle}{6} Q^2 + c Q^4 \, ,
  \label{eq:F_pi_qsqr_expansion}
\end{equation}
where the fit parameter $\langle r^2 \rangle$ is referred to as the \emph{charge radius} of the pion and the curvature $c$ is left unconstrained.
For the largest volume, at present, only the charge radius is included in the fit since there are only two points.
The results are tabulated in \cref{tab:F_pi_r_c_results}.

\renewcommand{\relfigwidth}{0.54}
\FPeval{\sidecaptionrelwidth}{1/\relfigwidth-1}
\begin{wrapfigure}{r}{\relfigwidth\textwidth}
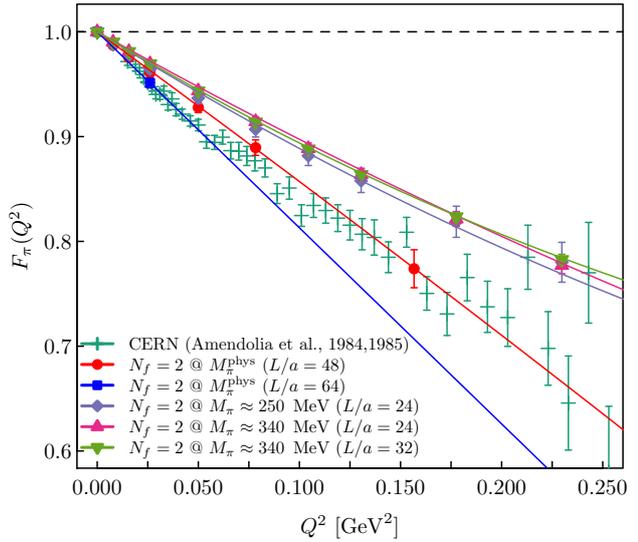

  \includegraphics[page=3,width=\relfigwidth\textwidth]{{{L64-L48-pionvectorff-gev}}}
  \caption{
    Vector pion form factor $F_\pi(Q^2)$ as a function of the space-like squared momentum transfer $Q^2$ at several pion masses and on several lattice volumes.
    The lines are fits of \cref{eq:F_pi_qsqr_expansion} to the data for each pion mass and volume, as given in \cref{tab:F_pi_r_c_results}. \\
  }
  \label{fig:etmc_pion_vectorff_gev}
\end{wrapfigure}

It is well known that the pion form factor (and consequently also the pion charge radius) are subject to significant finite size effects whether traditional lattice momenta or twisted boundary conditions are used.
As a result, the chiral extrapolation and infinite volume limits are intertwined in complicated ways.
It is common \cite{Frezzotti:2008dr,Brandt:2013dua}, therefore, to combine chiral perturbation ($\chi$-PT) theory fits and finite volume corrections for $F_\pi(Q^2)$ and  $\langle r^2 \rangle$ with corresponding fits of the pion mass and decay constant, thus extracting the common low energy constants (LECs) in the process.
Although our data reach down to the physical pion mass, it is unlikely that our determinations of $F_\pi(Q^2)$ and $\langle r^2 \rangle$ are free of finite size effects even on the largest volume.
Unfortunately, with presently only five ensembles, we are unable to perform the complete set of $\chi$-PT fits, such that we chose to follow an approach similar to the one taken in \cite{Koponen:2015tkr}.

\renewcommand{\relfigwidth}{0.6}
\FPeval{\sidecaptionrelwidth}{1/\relfigwidth-1}
\begin{SCfigure}
  \includegraphics[page=1,width=\relfigwidth\textwidth]{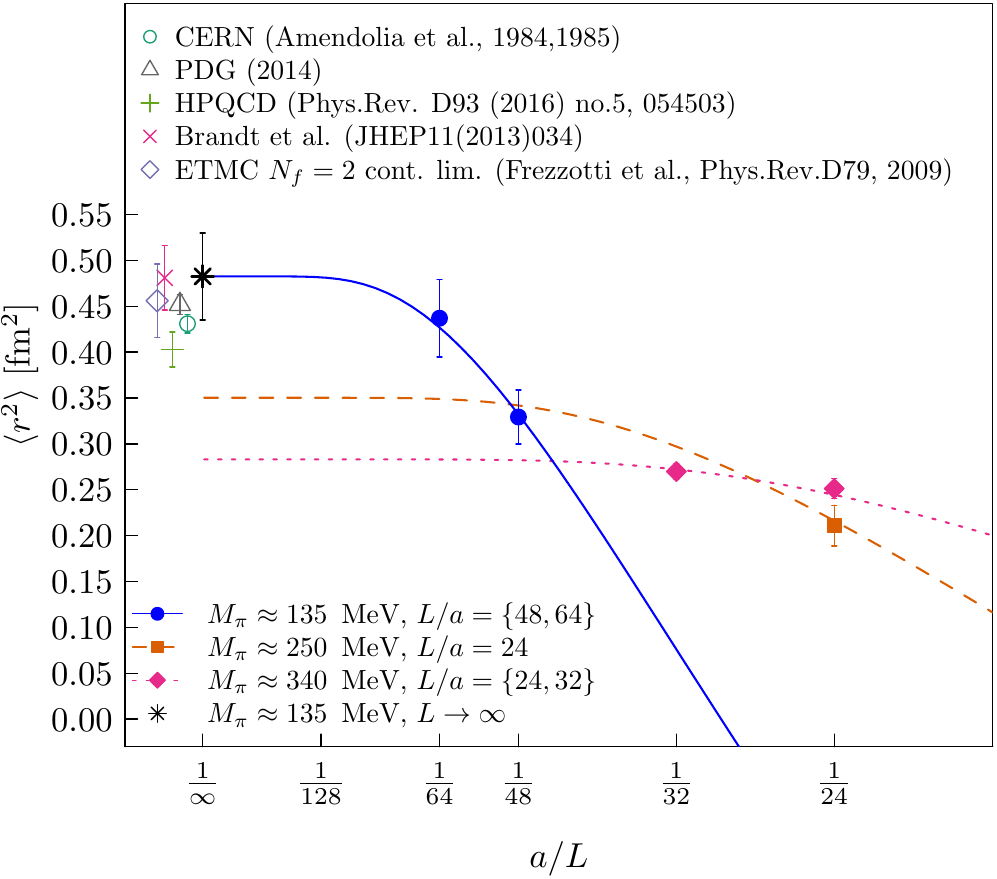}
  \caption{
    Infinite volume extrapolation following the Ansatz in \cref{eq:pionff_Vinf} of the squared pion charge radius $\langle r^2 \rangle$ based on data using four lattice volumes and three different pion masses, as given in \cref{tab:F_pi_r_c_results}.
  }
  \label{fig:etmc_pion_vectorff_volume_limit}
\end{SCfigure}

We carry out independent fits of $F_\pi(Q^2)$ at each pion mass and then perform an infinite volume extrapolation based on the following Ansatz which also includes the chiral logarithm:
\begin{equation}
  \left\langle r_L^2(\mpi) \right\rangle = \left( 1 - c_L\cdot\frac{\exp(-\mpi \cdot L)}{(\mpi \cdot L)^\alpha} \right) \cdot 
  \left[ \left\langle r^2_\infty(\mphyspi) \right\rangle - \frac{1}{\Lambda^2}\cdot\ln\left( \left[\frac{\mpi}{\mphyspi}\right]^2 \right) \right] \, ,
  \label{eq:pionff_Vinf}
\end{equation}
where  $\left\langle r^2_\infty(\mphyspi) \right\rangle$, $c_L$ and $\Lambda$ are taken as free fit parameters. 
Setting $\alpha = 3/2$, as for masses and decay constants, seems to give the best fit while fixing $\Lambda=\Lambda_\chi$ does not describe the data well at all, hinting at the presence of unquantified or higher order effects.
The result of the extrapolation is shown in \cref{fig:etmc_pion_vectorff_volume_limit}, giving
\begin{equation}
  \left\langle r^2_\infty(\mphyspi) \right\rangle = 0.46(4) \fm^2 \, .
\end{equation}
This is clearly compatible with the existing determinations from lattice QCD and phenomenology, although the final computation must improve upon \cref{eq:pionff_Vinf} to properly extrapolate to infinite volume and the continuum limit.

\section{Pion $\left\langle x \right\rangle$}

The method presented in \cref{sec:pionff} can also be used to compute $\langle x \rangle$, the quark momentum fraction of the pion, as well as higher moments of the quark momentum distribution.
At zero momentum transfer ($\theta_{1,2,3}=0$), the ratio of the pion two-point and three-point functions with the insertion  \cite{Best:1997qp}
\begin{equation*}
  \mathcal{O}_{v2b}(x) = \frac{1}{2} \bar{u}(x) \left( \gamma_4 \overleftrightarrow{D}_4 - \frac{1}{3} \sum_{i=1}^3 \gamma_i \overleftrightarrow{D}_i \right) u(x) \, ,
\end{equation*}
where $\overleftrightarrow{D}_\mu = \overrightarrow{D}_\mu - \overleftarrow{D}_\mu$, gives an estimate of $\langle x \rangle$ which is almost free of mixing under renormalisation.
It should be noted that we ignore the disconnected contribution to $\langle x \rangle$.

We evaluate these ratios on $N_f=2+1+1$ twisted mass ensembles \cite{Baron:2011sf} as well as the $N_f=2$ twisted mass ensembles with a clover term \cite{nf2physpointsimul:2015}. 
The results are shown in \cref{fig:etmc_pion_averx_overview}, with renormalisation constants computed in \cite{Alexandrou:2015sea}.
For the $N_f=2+1+1$ data, at each lattice spacing, the renormalised $\langle x \rangle$ are extrapolated to the physical pion mass quadratically in $M_\pi^\pm$.
These extrapolated points are shown in \cref{fig:etmc_pion_averx_final} together with a constant continuum extrapolation, which appears to be reasonably well justified within the present statistical and systematic uncertainties.
The final result
\begin{equation}
  \langle x \rangle_\mathrm{ren} = 0.246(05)(^{+11}_{-09}) \, ,
\end{equation}
is compatible with the phenomenological determination from \cite{Wijesooriya:2005ir}.
The data on the $N_f=2$ twisted mass ensembles with a clover term is compatible within errors, but appears to be systematically low compared to the other points, an effect which may well be explained by finite volume artefacts, as indicated by the $L=24$ and $L=32$ results at the highest pion mass in this set.

\renewcommand{\relfigwidth}{0.63}
\FPeval{\sidecaptionrelwidth}{1/\relfigwidth-1}
\begin{SCfigure}
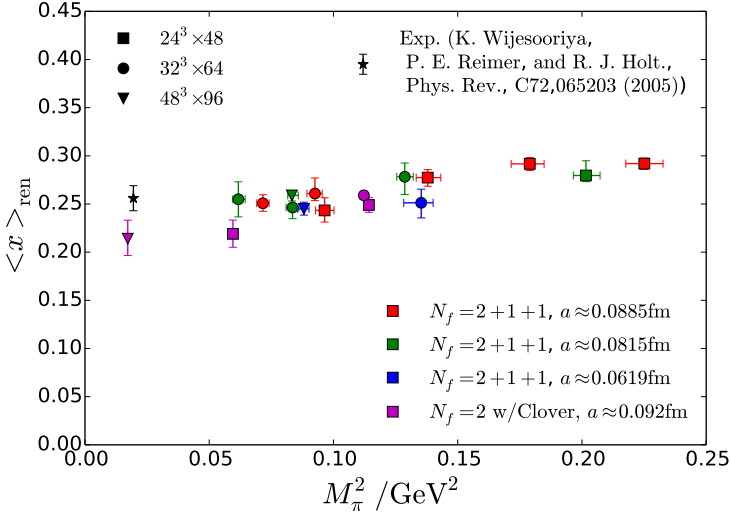

  \centering
  \vspace{0.5cm} \newline
  \includegraphics[width=\relfigwidth\textwidth]{{{etmc_pion_averx_overview}}}
  \caption{Determination of the renormalised pion $\langle x \rangle$ on $N_f=2+1+1$ twisted mass ensembles at three lattice spacings with pion masses down to about $240\, \mathrm{MeV}$ and on $N_f=2$ twisted mass ensembles with a clover term including a physical pion mass. The phenomenological is taken from \cite{Wijesooriya:2005ir}.}
  \label{fig:etmc_pion_averx_overview}
\end{SCfigure}

\begin{SCfigure}
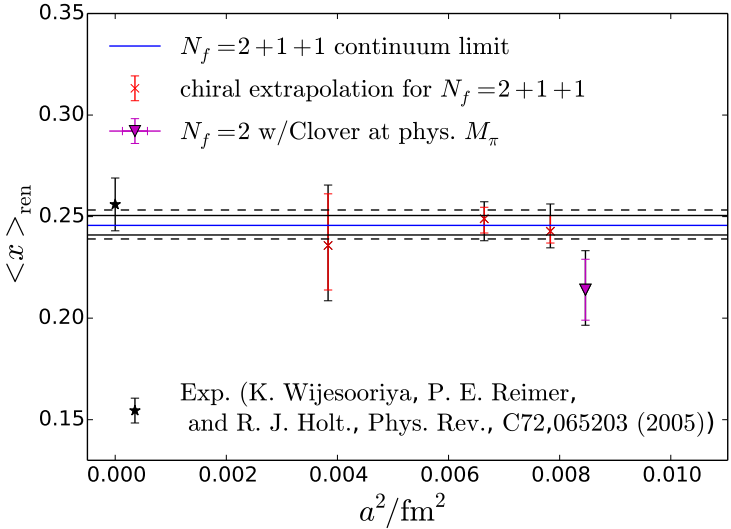

  \centering
  \vspace{0.5cm} \newline
  \includegraphics[width=\relfigwidth\textwidth]{{{etmc_pion_averx_final}}}
  \caption{Constant continuum extrapolation of the renormalised $\left\langle x \right\rangle$ of the pion as computed on $N_f=2+1+1$ twisted mass ensembles, chirally extrapolated to $M_\pi^\mathrm{phys}$ at each value of the lattice spacing. For comparison, a determination on an $N_f=2$ twisted mass ensemble with a clover term at the physical pion mass is given, as well as the phenomenological determination from \cite{Wijesooriya:2005ir}. The continuum limit result is indicated by the horizontal line. The inner and outer error bars on the individual points as well as the final result indicate the unsummed statistical and systematic errors respectively.}
  \label{fig:etmc_pion_averx_final}
\end{SCfigure}

\section{Conclusions and Outlook}

We have presented computations of the electromagnetic form factor as well as the connected contribution to the quark momentum fraction of the pion.
Both of these have been shown to be largely compatible with phenomenological as well as other lattice determinations.

We are in the process of extending both calculations.
For the pion form factor, significantly increased statistics, additional momentum transfers as well as the analysis of a further ensemble of gauge configurations will provide a data set which will hopefully allow us to control most systematic errors in our computation except for discretisation effects.
We are extending our study of the moments of distribution functions of the pion to a statistically significant determination of the third moment, which should allow us to compute a value for $\left\langle x^2 \right\rangle$ in the continuum limit.
In this process, we should be able to also increase the precision of our determination of $\left\langle x \right\rangle$.

\section{Acknowledgements}

We would like to thank the members of the ETMC for the most enjoyable collaboration.
The computing time for this project has been made available by the John von Neumann-Institute for Computing (NIC) on the Jureca and Juqueen systems in J{\"u}lich.
This project was funded by the DFG as a part of the Sino-German CRC110.

\bibliographystyle{h-physrev5}

\begin{thebibliography}{10}

\bibitem{Frezzotti:2003ni}
R.~Frezzotti and G.~C. Rossi,
\newblock JHEP {\bf 08}, 007 (2004),
  \href{http://arxiv.org/abs/hep-lat/0306014}{{\tt arXiv:hep-lat/0306014
  [hep-lat]}}.

\bibitem{Frezzotti:2008dr}
{\bf ETM} Collaboration, R.~Frezzotti, V.~Lubicz and S.~Simula,
\newblock Phys. Rev. {\bf D79}, 074506 (2009),
  \href{http://arxiv.org/abs/0812.4042}{{\tt arXiv:0812.4042 [hep-lat]}}.

\bibitem{Boyle:2008yd}
P.~A. Boyle {\em et~al.},
\newblock JHEP {\bf 07}, 112 (2008), \href{http://arxiv.org/abs/0804.3971}{{\tt
  arXiv:0804.3971 [hep-lat]}}.

\bibitem{Brandt:2013dua}
B.~B. Brandt, A.~J{\"u}ttner and H.~Wittig,
\newblock JHEP {\bf 11}, 034 (2013), \href{http://arxiv.org/abs/1306.2916}{{\tt
  arXiv:1306.2916 [hep-lat]}}.

\bibitem{Koponen:2015tkr}
J.~Koponen, F.~Bursa, C.~T.~H. Davies, R.~J. Dowdall and G.~P. Lepage,
\newblock Phys. Rev. {\bf D93}, 054503 (2016),
  \href{http://arxiv.org/abs/1511.07382}{{\tt arXiv:1511.07382 [hep-lat]}}.

\bibitem{Blok:2008jy}
{\bf Jefferson Lab} Collaboration, H.~P. Blok {\em et~al.},
\newblock Phys. Rev. {\bf C78}, 045202 (2008),
  \href{http://arxiv.org/abs/0809.3161}{{\tt arXiv:0809.3161 [nucl-ex]}}.

\bibitem{Huber:2008id}
{\bf Jefferson Lab} Collaboration, G.~M. Huber {\em et~al.},
\newblock Phys. Rev. {\bf C78}, 045203 (2008),
  \href{http://arxiv.org/abs/0809.3052}{{\tt arXiv:0809.3052 [nucl-ex]}}.

\bibitem{Amendolia:1986wj}
{\bf NA7} Collaboration, S.~R. Amendolia {\em et~al.},
\newblock Nucl. Phys. {\bf B277}, 168 (1986).

\bibitem{deDivitiis:2004kq}
G.~M. de~Divitiis, R.~Petronzio and N.~Tantalo,
\newblock Phys. Lett. {\bf B595}, 408 (2004),
  \href{http://arxiv.org/abs/hep-lat/0405002}{{\tt arXiv:hep-lat/0405002
  [hep-lat]}}.

\bibitem{Sachrajda:2004mi}
C.~T. Sachrajda and G.~Villadoro,
\newblock Phys. Lett. {\bf B609}, 73 (2005),
  \href{http://arxiv.org/abs/hep-lat/0411033}{{\tt arXiv:hep-lat/0411033
  [hep-lat]}}.

\bibitem{Flynn:2005qn}
J.~Flynn, A.~Juttner, C.~Sachrajda and G.~Villadoro,
\newblock PoS {\bf LAT2005}, 352 (2006),
  \href{http://arxiv.org/abs/hep-lat/0509093}{{\tt arXiv:hep-lat/0509093
  [hep-lat]}}.

\bibitem{nf2physpointsimul:2015}
{\bf ETM} Collaboration, A.~Abdel-Rehim {\em et~al.},
\newblock \href{http://arxiv.org/abs/1507.05068}{{\tt arXiv:1507.05068
  [hep-lat]}}.

\bibitem{Best:1997qp}
C.~Best {\em et~al.},
\newblock Phys. Rev. {\bf D56}, 2743 (1997),
  \href{http://arxiv.org/abs/hep-lat/9703014}{{\tt arXiv:hep-lat/9703014
  [hep-lat]}}.

\bibitem{Baron:2011sf}
{\bf ETM} Collaboration, R.~Baron {\em et~al.},
\newblock PoS {\bf LATTICE2010}, 123 (2010),
  \href{http://arxiv.org/abs/1101.0518}{{\tt arXiv:1101.0518 [hep-lat]}}.

\bibitem{Alexandrou:2015sea}
{\bf ETM} Collaboration, C.~Alexandrou, M.~Constantinou and H.~Panagopoulos,
\newblock \href{http://arxiv.org/abs/1509.00213}{{\tt arXiv:1509.00213
  [hep-lat]}}.

\bibitem{Wijesooriya:2005ir}
K.~Wijesooriya, P.~E. Reimer and R.~J. Holt,
\newblock Phys. Rev. {\bf C72}, 065203 (2005),
  \href{http://arxiv.org/abs/nucl-ex/0509012}{{\tt arXiv:nucl-ex/0509012
  [nucl-ex]}}.

\end{thebibliography}

\end{document}